\documentclass[conference]{IEEEtran}
\IEEEoverridecommandlockouts
\usepackage{cite}
\usepackage{amsmath,amssymb,amsfonts}
\usepackage{algorithmic}
\usepackage{graphicx}
\usepackage{textcomp}
\usepackage{xcolor}
\def\BibTeX{{\rm B\kern-.05em{\sc i\kern-.025em b}\kern-.08em
    T\kern-.1667em\lower.7ex\hbox{E}\kern-.125emX}}

\usepackage{caption}
\usepackage{subcaption}
\usepackage{multirow}
\usepackage{enumitem}

\newcommand{\Ls}{\mathcal{L}}
\def\sP{{\mathbb{P}}}

\DeclareMathOperator*{\argmax}{arg\,max}

\begin{document}

\title{SafeAMC: Adversarial training for robust modulation recognition models
\thanks{This work has been sponsored by armasuisse Science and Technology with the project ROBIN (project code Aramis 047-22).}
}

\author{\IEEEauthorblockN{Javier Maroto\IEEEauthorrefmark{1},
Gérôme Bovet\IEEEauthorrefmark{2} and
Pascal Frossard\IEEEauthorrefmark{1}}
\IEEEauthorblockA{\IEEEauthorrefmark{1}EPFL, Switzerland}
\IEEEauthorblockA{\IEEEauthorrefmark{2}armasuisse Science\&Technology, Cyber-Defence Campus, Switzerland}
}

\maketitle

\begin{abstract}
In communication systems, there are many tasks, like modulation recognition, which rely on Deep Neural Networks (DNNs) models. However, these models have been shown to be susceptible to adversarial perturbations, namely imperceptible additive noise crafted to induce misclassification. This raises questions about the security but also the general trust in model predictions. We propose to use adversarial training, which consists of fine-tuning the model with adversarial perturbations, to increase the robustness of automatic modulation recognition (AMC) models. We show that current state-of-the-art models benefit from adversarial training, which mitigates the robustness issues for some families of modulations. We use adversarial perturbations to visualize the features learned, and we found that in robust models the signal symbols are shifted towards the nearest classes in constellation space, like maximum likelihood methods. This confirms that robust models not only are more secure, but also more interpretable, building their decisions on signal statistics that are relevant to modulation recognition.
\end{abstract}


\section{Introduction}

Communication systems are important for both civil and military applications. Deep learning \cite{goodfellow2016deep} not only proved its usefulness across multiple fields of research in the last decade but presents numerous advantages that makes it attractive for wireless communication systems too. Compared with the previous state-of-the-art approaches, which are mainly based on feature extraction from the signals \cite{Dobre_Abdi_Bar-Ness_Su_2007}, Deep Neural Networks (DNNs) scale well with high quantities of data and are capable of end-to-end learning, which simplifies the model architecture and improves performance. DNNs have been successful in multiple tasks like wireless resource allocation \cite{Sun_Chen_Shi_Hong_Fu_Sidiropoulos_2017} anomaly detection \cite{Chalapathy_Chawla_2019}, or automatic modulation recognition (AMC) \cite{OShea_Roy_Clancy_2018}, which is the main focus of this work.
Modulation recognition is required to interpret the received data and allows for multiple applications ranging from detecting daily radio stations and managing spectrum resources, to eavesdropping and interfering with radio communications.

However, recent studies have highlighted security issues in DNNs models \cite{Szegedy_Zaremba_Sutskever_Bruna_Erhan_Goodfellow_Fergus_2014, Moosavi-Dezfooli_Fawzi_Fawzi_Frossard_2017,Sadeghi_Larsson_2019,Lin_Zhao_2020,Flowers_Buehrer_Headley_2019, maroto2021benefits}. Specifically, they have been shown to be vulnerable to adversarial examples, which add a carefully crafted but almost imperceptible perturbation, namely adversarial perturbation, to a real data sample. This observation combined with the black-box nature of DNNs raises a question: can we trust the predictions of neural networks? The fact that a negligible change in the input would change the prediction, implies that DNNs base their decisions on features that do not seem to be aligned with the target task. Understanding the reasons for such vulnerabilities and making systems more robust form an active line of research.


The main contributions we provide in this paper are the followings:
\begin{itemize}[leftmargin=*,nosep]
    \item We are the first to propose a defense mechanism against adversarial examples in AMC, namely adversarial training, which successfully increases the robustness of these models;
    \item We design a new framework based on the specific properties of communication systems. It better captures and measures how secure AMC models are to adversarial attacks, inspired by realistic settings. Thus, we are the first to differentiate between practical security and robustness in AMC;
    \item We do an extensive analysis of the robustness and security of some state-of-the-art AMC models on some popular modulation recognition datasets. Using feature analysis on the constellation diagram space, we show that robust models learn features that are more aligned with the optimal ones learned by Bayes-optimal maximum likelihood methods;
    \item Finally, we promote good practices when tackling robustness on AMC (e.g., constraining the perturbation relative to the signal energy).
\end{itemize}
In Section~\ref{sec:related_work} we discuss some related works related to AMC and adversarial perturbations. Then, in Section~\ref{sec:framework}, we introduce our adversarial learning framework with two dedicated use cases, namely robustness and security. We provide results of the experiments realized along with our newly defined framework in Section~\ref{sec:experiments}. The feature importance of standardly and adversarially trained models is presented in Section~\ref{sec:features}. We finally conclude and provide future work directions in Section~\ref{sec:conclusion}.

\section{Related work}
\label{sec:related_work}

The task of recognizing modulations can be modeled mathematically using maximum likelihood. Maximum likelihood methods \cite{Chung-Yu_Huan_Polydoros_1995,Dobre_Abdi_Bar-Ness_Su_2007,Hameed_Dobre_Popescu_2009} approximate the likelihood function of the received signal with all possible modulations and estimate the most likely modulation. It requires prior knowledge of the channel characteristics, which makes its use difficult in the real world, where unexpected channel corruptions could happen. However, if priors are correct, then the exact computation of the likelihood function is Bayes-optimal, which leads to perfect modulation recognition. However, in practice, computing the exact likelihood function is unfeasible and sub-optimal approximations are used instead.

Recently, deep learning \cite{goodfellow2016deep} has been proposed as a better solution since it has low complexity which, unlike maximum likelihood methods, scales linearly, and can be trained end-to-end. This allows for extra flexibility, since it does not require any assumption on the channel conditions but learns them directly from the data. Two types of Deep Learning models have been proposed for modulation recognition. On the one hand, inspired by successes in speech recognition, some works \cite{Rajendran_Meert_Giustiniano_Lenders_Pollin_2018,Guo_Jiang_Wu_Zhou_2020} propose architectures based on long short-term memory networks (LSTMs) \cite{Hochreiter_Schmidhuber_1997} for modulation recognition. On the other hand, other works \cite{OShea_Corgan_Clancy_2016,West_OShea_2017,Sadeghi_Larsson_2019} employ convolutional neural networks (CNNs) \cite{Krizhevsky_Sutskever_Hinton_2017}, which have the advantage of inducing translation invariance on the input. The current state-of-the-art in modulation recognition \cite{OShea_Roy_Clancy_2018} uses a model based on the ResNet Deep Network architecture \cite{Szegedy_Ioffe_Vanhoucke_Alemi_2016}. 

Despite their recent popularity and good performance in various applications, these DNNs have some drawbacks. In particular, they are black-box models, making their predictions much more difficult to explain than their classical counterparts. They are also vulnerable to imperceptible crafted noise added to the input, forming what is commonly referred to as adversarial perturbations \cite{Szegedy_Zaremba_Sutskever_Bruna_Erhan_Goodfellow_Fergus_2014, Moosavi-Dezfooli_Fawzi_Fawzi_Frossard_2017}. These perturbations have the property of influencing discriminative features of the model \cite{Engstrom_Ilyas_Santurkar_Tsipras_Tran_Madry_2019}, and they greatly affect the model prediction. Since they are imperceptible or very small, they put into question the relevance of the features learned by the classifier \cite{ilyas2019adversarial}. To improve the model robustness against adversarial perturbations, multiple works propose different approaches or ``defenses" in different application domains. Some authors advocate for randomized smoothing \cite{Cohen_Rosenfeld_Kolter_2019, Salman_Li_Razenshteyn_Zhang_Zhang_Bubeck_Yang_2019}, while others propose the addition of some kind of regularization to the loss function \cite{Moosavi-Dezfooli_Fawzi_Uesato_Frossard_2019, Jagatap_Chowdhury_Garg_Hegde_2020}. Currently, the best defense relies on adversarial training \cite{Madry_Makelov_Schmidt_Tsipras_Vladu_2019}, where the model is fine-tuned on adversarial perturbations in order to increase its robustness.

In the case of modulation recognition, adversarial perturbations have also been shown to be effective and to require much less power than additive white gaussian noise (AWGN) to fool the network \cite{maroto2021benefits,Sadeghi_Larsson_2019,Flowers_Buehrer_Headley_2019}. Adversarial perturbations in this case are constrained relatively to the signal power, using the signal-to-perturbation ratio (SPR) metric \cite{Sadeghi_Larsson_2019}. However, little work has been done to defend against perturbations and make modulation recognition more robust.


\section{Framework}
\label{sec:framework}

In this work, we present SafeAMC, a new framework based on adversarial training to make DNNs less susceptible to adversarial perturbations. We show that it increases the robustness of these models, making them safer to use.

\subsection{Adversarial Perturbations}

The objective of adversarial perturbations is to minimally modify the input signal such that the model outputs wrong predictions. The perturbation should be `imperceptible', that is, it is so small that they do not change the true label of the classification task. For modulation recognition, we are mostly interested in perturbations with small $l_{\infty}$ norm. Thus, we define adversarial perturbations as:
\begin{equation}
\label{eq:adv_pert}
    \delta_i^* = \argmax_{\delta_i}\Ls_{y_i, \theta}(x_i + \delta_i) \quad \text{s.t.} \quad \lVert \delta_i \rVert_{\infty} \leq \varepsilon
\end{equation}
where $\delta_i^*$ is the adversarial perturbation, $x_i$ is the clean signal, $y_i$ is the true label, $\lVert \cdot \rVert_{\infty}$ is the $l_{\infty}$ norm, $\theta$ are the model parameters, $\Ls_{y_i, \theta}$ is the model loss function, and $\varepsilon$ is the constraint imposed on the norm of the adversarial perturbation.

The maximization problem that must be solved to obtain the adversarial perturbation is a difficult optimization problem of its own. Some of the most popular solutions use the gradient with respect to the input to find a good approximation to the maximizer in Eq \eqref{eq:adv_pert}. Iterative algorithms like Projected Gradient Descent (PGA)\footnote{The PGA algorithm is commonly referred in the adversarial perturbation literature as Projected Gradient Descent (PGD), which is a misnomer given its definition.} \cite{Madry_Makelov_Schmidt_Tsipras_Vladu_2019} are expensive since they require multiple backpropagation steps, while one-step algorithms like Fast Gradient Sign Method (FGSM) only requires one backpropagation pass but the adversarial perturbation is weaker. We write the objectives of the FGSM and PGA algorithms in equations \eqref{eq:fgsm} and \eqref{eq:pga}, respectively:
\begin{align}
    \label{eq:fgsm} \delta_i^* &\approx \varepsilon \text{sign}(\nabla_x \Ls_{y_i, \theta}(x_i)) \\
    \label{eq:pga} \delta_i^* &\approx \delta_i^{K} \qquad \delta_i^{k+1} = \sP_{\mathcal{S}_i}(\delta_i^{k} + \beta \varepsilon \nabla_x \Ls_{y_i, \theta}(x_i + \delta_i^{k}))
\end{align}
where $\nabla_x$ is the gradient with respect to the input space, $\mathcal{S}_i$ is the region inside the $l_{\infty}$ ball of radius $\varepsilon$ centered at the point $x_i$, $\beta$ is the step size of PGA, $K$ is the number of iterations of PGA, and $\sP_{\mathcal{S}_i}$ is the projection operator onto the region $\mathcal{S}_i$.

\subsection{Adversarial training}

The objective of adversarial training is to find the optimal model parameters that minimize the susceptibility of the model to adversarial perturbations. That is, adversarial training tries to solve the following min-max optimization problem over the $N$ training samples:
\begin{equation}
\label{eq:emp_adv_risk}
    \min_{\theta} \dfrac{1}{N} \sum_i \max_{\lVert \delta_i \rVert_{\infty} \leq \varepsilon} \Ls_{y_i, \theta}(x_i + \delta_i)
\end{equation}
where the inner-maximization objective is to find the adversarial perturbation, and the outer-minimization tries to find the model parameters that minimize the impact of adversarial perturbations in the classification loss. This formulation can be seen as a generalization of standard training, which rather finds the optimal model parameters when there are no adversarial perturbations. That is, $\delta_i = 0$, which effectively removes the inner-maximization.

In practice, the procedure to train a model adversarially is straightforward. First, we take a training sample and compute the adversarial perturbation using FGSM or PGA. Then, we compute the model classification loss on that perturbed sample. Next, we backpropagate this loss with respect to the model parameters. Finally, we update them to minimize the loss using gradient descent. This whole process is repeated for all training samples multiple times. Since the model parameters change on every training iteration, the adversarial perturbations cannot be reused between different epochs. This makes adversarial training slow when compared with standard training. Improving the convergence speed of the min-max optimization of the empirical adversarial risk is an active line of research.

\subsection{Robustness evaluation in modulation recognition}

\begin{figure}
    \centering
    \begin{subfigure}[b]{0.7\linewidth}
        \centering
        \includegraphics[width=\textwidth]{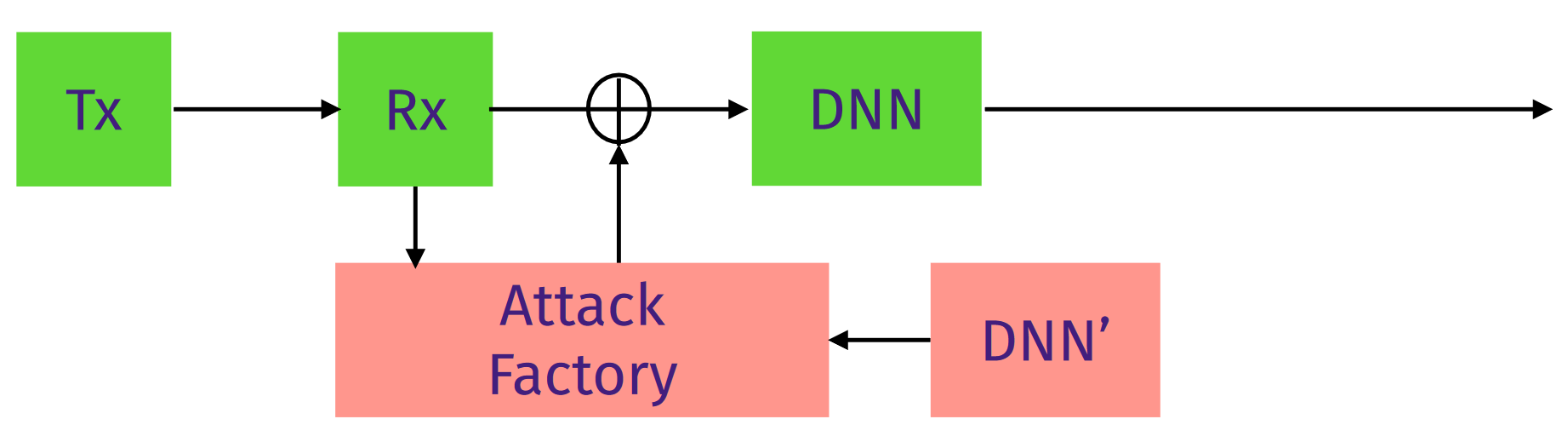}
        \caption{Robustness framework}
        \label{fig:rob_frwk}
    \end{subfigure}
    \\[2ex]
    \begin{subfigure}[b]{0.7\linewidth}
        \centering
        \includegraphics[width=\textwidth]{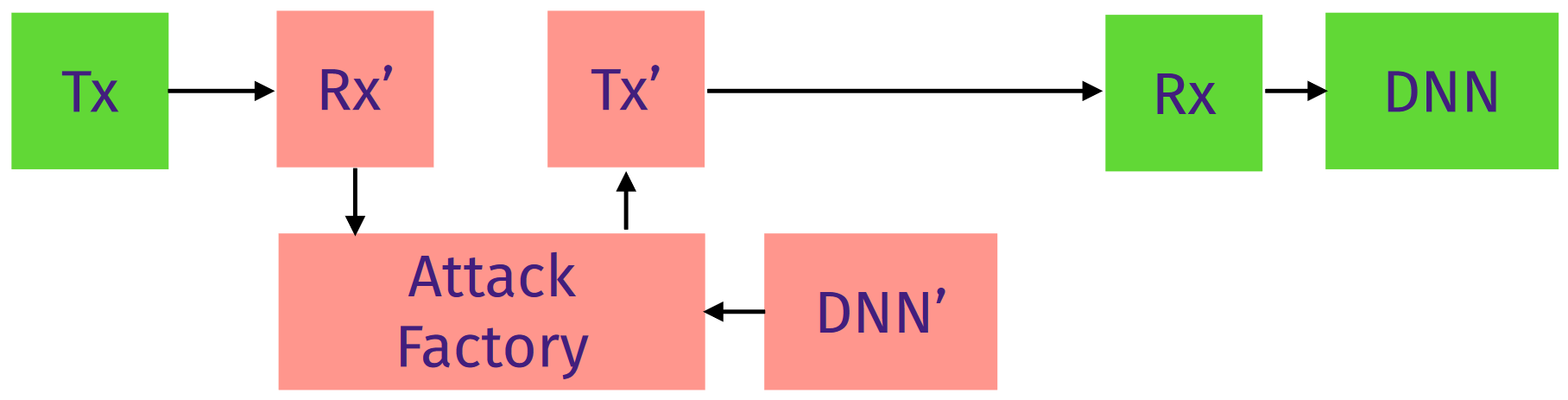}
        \caption{Security framework}
        \label{fig:sec_frwk}
    \end{subfigure}
    \caption{Frameworks used to measure robustness and security. The green and red blocks illustrate the communication system and the attacker, respectively. Tx and Rx are the transmitter and receiver, while DNN is the modulation recognition model. The attacker uses a surrogate model to craft an adversarial perturbation but, since we are analyzing the worst case scenario, we use DNN = DNN' (white-box attack).}
    \label{fig:frameworks}
\end{figure}

In modulation recognition, to have a realistic measure of how secure the model is against malicious attacks, the framework has to take into account that communication corruptions may be added between the attacker and the receiver side. This motivates us to use two different frameworks to measure robustness and security \cite{maroto2021benefits}, which are shown in Figure \ref{fig:frameworks}. The robustness framework is similar to the traditional approach, adding the attack just before the model. We use it to measure the theoretical robustness of our model. The security framework makes a more realistic assumption in which the attacker is in the middle of the communication system. Channel corruptions, modeled as Gaussian noise, are added to the adversarial example transmitted by the attacker. Thus, providing a better measure on how secure the model is against adversarial attacks.

Our framework, SafeAMC, trains the model using the robustness framework to help the model learn signal features that are invariant to small perturbations. Then, it evaluates the model on both frameworks to measure the robustness and the security of the model.

\section{Experiments}
\label{sec:experiments}
\subsection{Datasets}

We analyze the robustness and security of the state-of-the-art models when using adversarial training, as proposed in the previous Section. We use the widely used RML2016.10a \cite{OShea_West_2016} and RML2018.01a \cite{OShea_Roy_Clancy_2018} DeepSig RadioML datasets for this analysis. The RML2016.10a dataset has 220000 IQ signals of 128 time samples each, ranging from -20dB to 18dB signal-to-noise ratio (SNR), with 11 possible modulations. The RML2018.01a dataset presents a harder task. It contains more than 2.5 million IQ signals of 1024 time samples each, ranging from -20dB to 30dB SNR, with 24 possible modulations.

We split the RML2016.10a IQ signals into 70\% of training and 30\% of test signals, using 5\% of the training as validation for hyperparameter tuning, as proposed by \cite{Flowers_Buehrer_Headley_2019}. For the RML2018.01a dataset, we use 1 million signals as training set as proposed in \cite{OShea_Roy_Clancy_2018}. From the remaining ones we use 5\% for validation, and the rest for the test set.

Additionally, for the feature analysis we describe in the next Section, we created a custom simulated dataset (CRML2018) where we have knowledge of the symbols transmitted and have full control on the communication system parameters. We generated 10000 signals of 1024 time samples for each of the 16 digital modulations used in the RML2018.01a dataset. We corrupt all of them with 20 dB SNR AWGN, making the task much simpler to learn since there are no noisy signals. We use 70\% of the data for training and 30\% for testing, using 5\% of the training as validation for hyperparameter tuning.

\subsection{Performance analysis}

First, we do a small experiment on the RML2016.10a dataset, to see how adversarial training changes the model predictions on adversarial examples. We use the robustness framework described before with the VT\_CNN2\_BF architecture \cite{Flowers_Buehrer_Headley_2019} as our AMC model. For both the training and testing adversarial examples, we choose 20 dB for the SPR since it decreases significantly the performance of the model without compromising in perceptibility \cite{maroto2021benefits}. We use $l_{\infty}$ PGA with 7 iterations of step size 0.36 (relative to the $l_{\infty}$ ball radius) for the training adversarial examples, and PGA with 20 iterations of step size 0.125 for the testing adversarial examples.

\begin{figure}[tbp]
    \centering
    \begin{subfigure}[b]{0.45\linewidth}
        \centering
        \includegraphics[width=\textwidth]{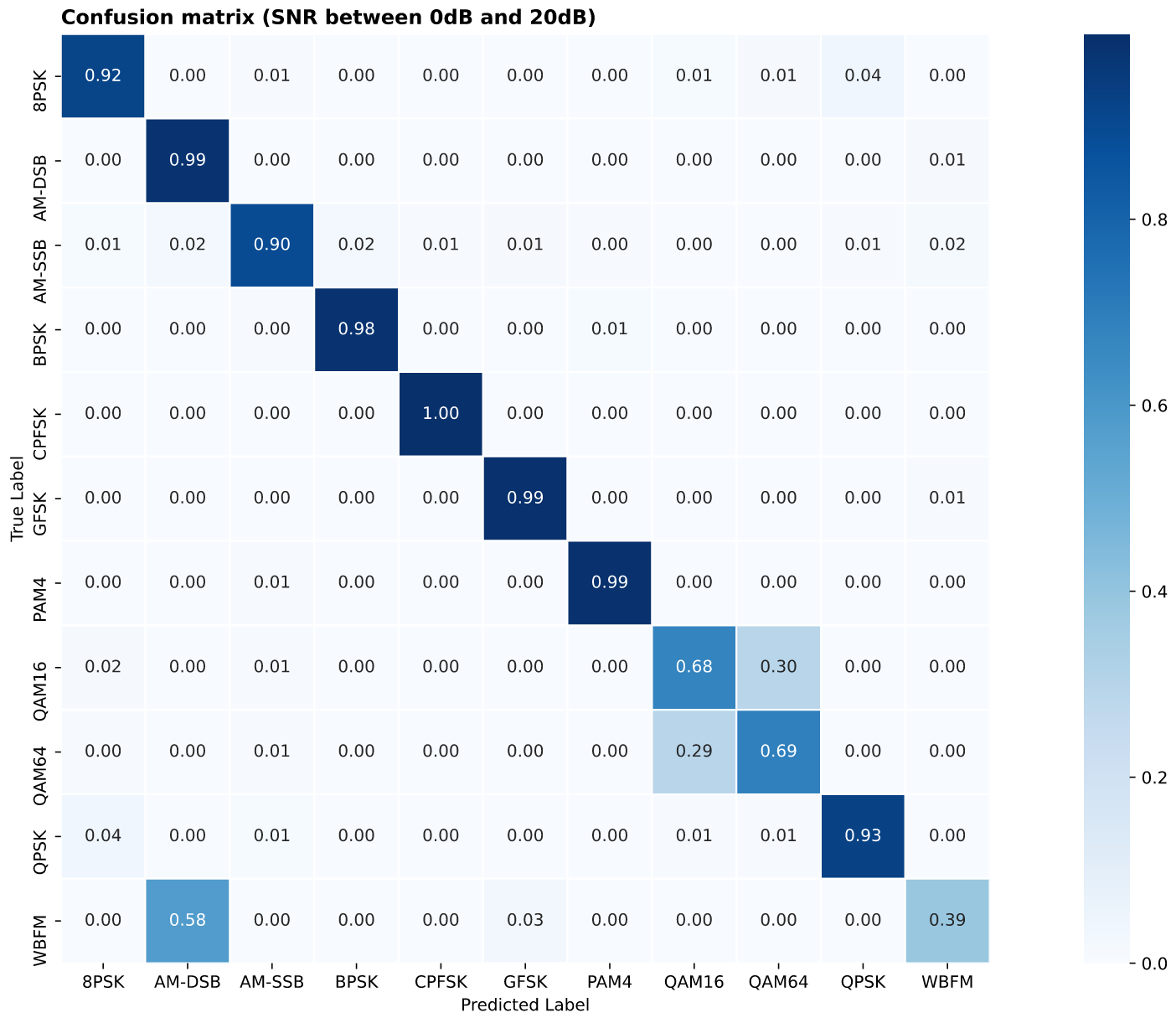}
    \end{subfigure}
    \;
    \begin{subfigure}[b]{0.45\linewidth}
        \centering
        \includegraphics[width=\textwidth]{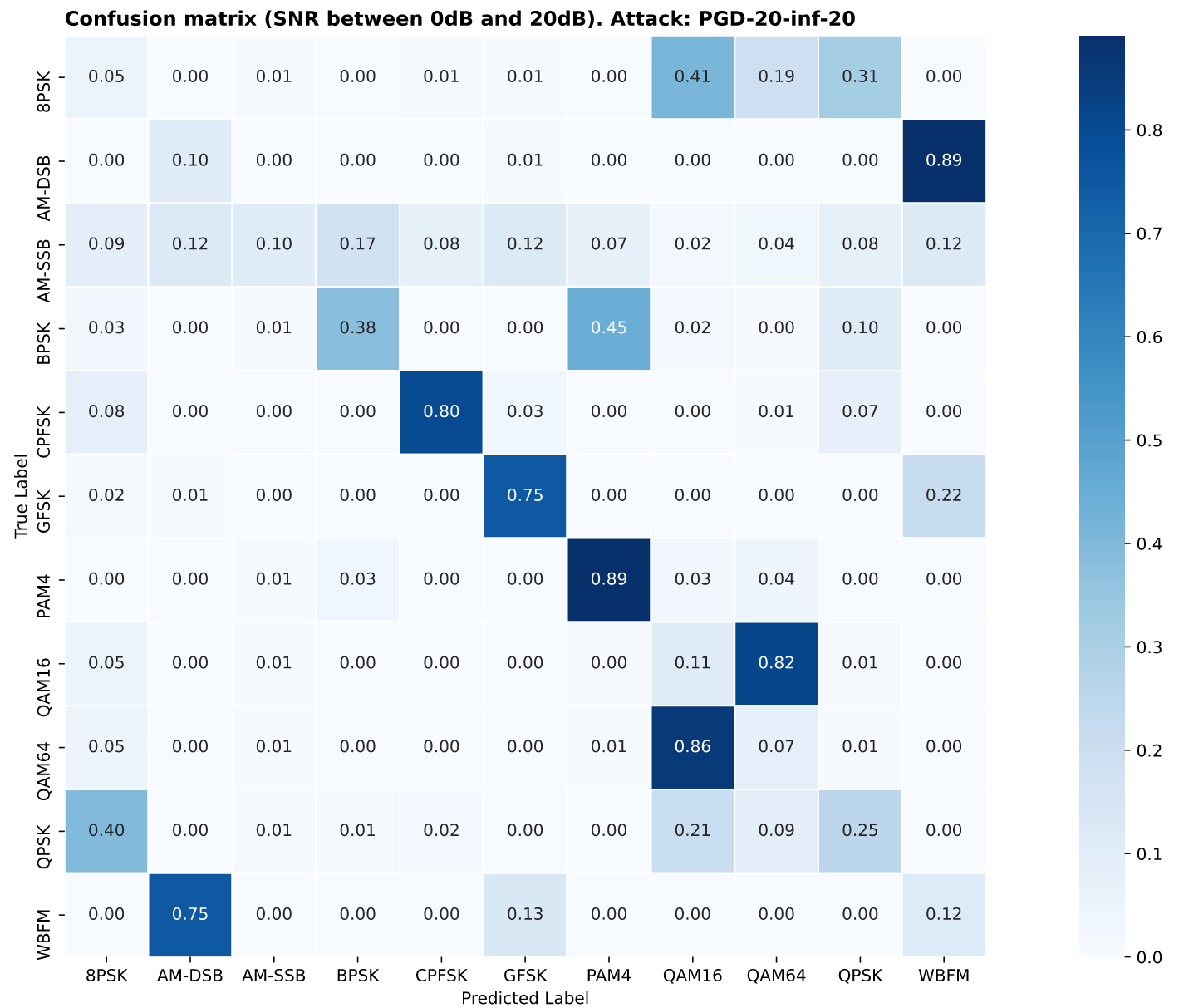}
    \end{subfigure}
    \caption{Confusion matrices of the VT\_CNN2\_BF model before (left) and after (right) adding PGA adversarial perturbations. Results for IQ signals of SNR higher or equal than 0dB.}
    \label{fig:cm_std}
\end{figure}

\begin{figure}[tbp]
    \centering
    \begin{subfigure}[b]{0.45\linewidth}
        \centering
        \includegraphics[width=\textwidth]{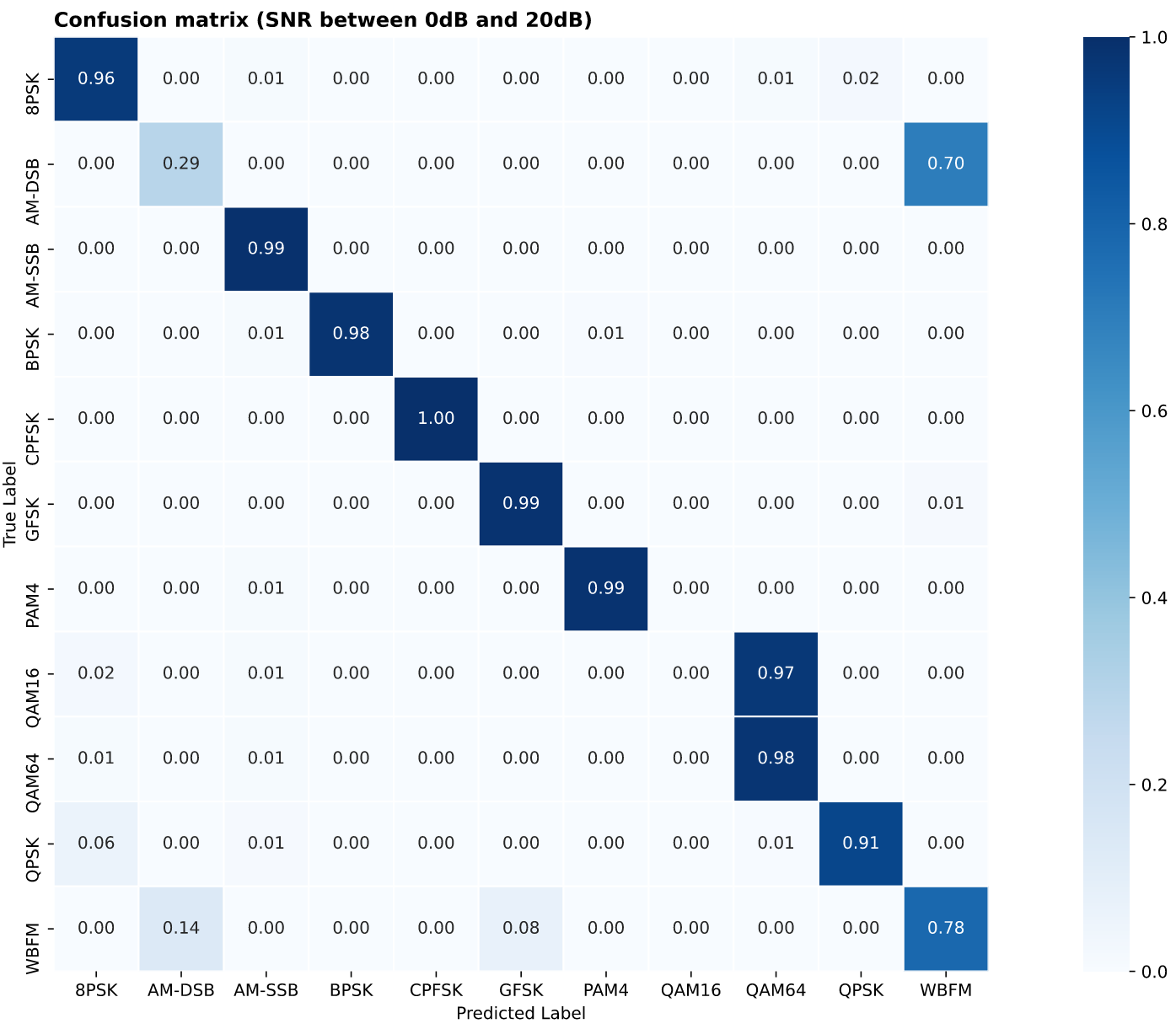}
    \end{subfigure}
    \;
    \begin{subfigure}[b]{0.45\linewidth}
        \centering
        \includegraphics[width=\textwidth]{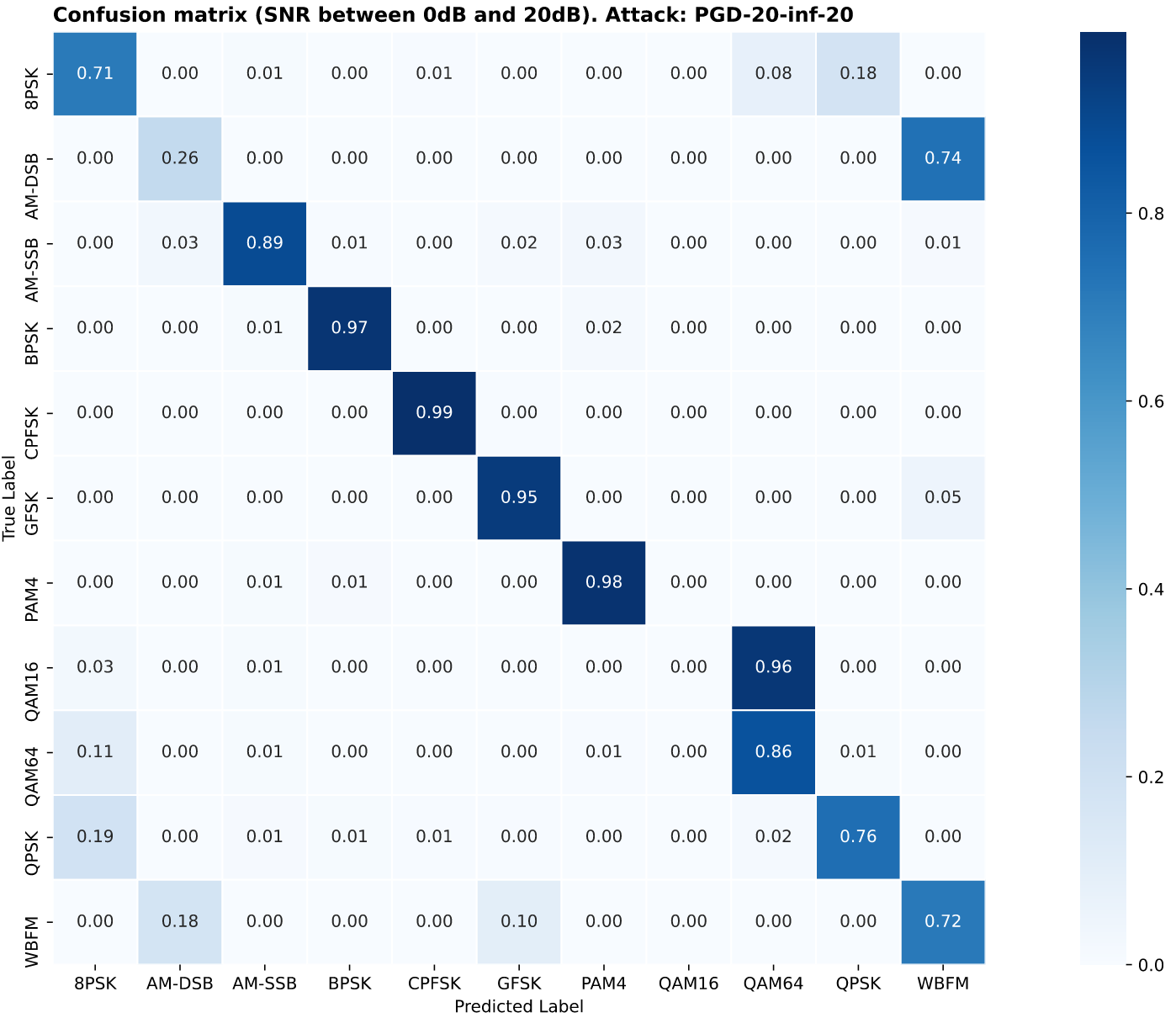}
    \end{subfigure}
    \caption{Confusion matrices of the adversarially trained VT\_CNN2\_BF model before (left) and after (right) adding PGA adversarial perturbations. Results for IQ signals of SNR higher or equal than 0dB.}
    \label{fig:cm_adv}
\end{figure}

We show the results of this first analysis on Figures \ref{fig:cm_std} and \ref{fig:cm_adv}, revealing the confusion matrix on the original and the perturbed test data of the standardly and the adversarially trained AMC model, respectively. If we focus on the standardly trained model, we can see that the model performs significantly worse when attacked. There are some classes where the model is robust (e.g., PAM4), but generally the model is fooled towards modulations with similar constellation diagrams (e.g. BPSK towards PAM4, and QAM16 towards QAM64). When that same model is trained adversarially instead, we show that it is significantly more robust against adversarial perturbations, but at the cost of some performance when tested on the original data. We highlight two pairs of modulations where the performance suffers the most: AM-DSB/WBFM, in which due to dataset generation errors, the first of them consists of Gaussian noise signals; and QAM16/QAM64. For the latter pair, the model seems unable to differentiate between both classes after being adversarially trained, behaving like a constant classifier. We believe the SPR of the perturbation is too low in this case, changing the signal class between these two high-order modulations.

For our next experiment, we benchmark the robustness as well as the security of two state-of-the-art DNN models, the VT\_CNN2\_BF and ResNet models, on the more difficult RML2018.01a dataset. For this purpose, we use SafeAMC. For the security framework, we add 20 dB SNR AWGN to the signal transmitted by the attacker. Like in the previous experiment, we use $l_{\infty}$ PGA with 7 iterations of step size 0.36 for the training adversarial examples, and PGA with 20 iterations of step size 0.125 for the testing adversarial examples. However, this time we compare between different SPR values for both the training and testing adversarial perturbations. We consider standard training and adversarial training using 15 dB and 20 dB for the SPR. For the testing, we consider 15 dB, 20 dB, and 25 dB SPR as well as no perturbation at all. The comparison of the accuracy of the AMC models on all the training schemes, different testing scenarios, and different frameworks used is shown in Table \ref{tab:vt_cnn} (VT\_CNN2\_BF model) and Table \ref{tab:resnet} (ResNet model).

The results of this more extensive analysis show some clear patterns. First, we can evaluate the effect of the AWGN added to the security framework. For standardly trained models, this addition reduces the fooling rate of the attack compared to the robustness framework. Thus making the attacker efforts less effective when it does not have access to the receiver itself. However, if the model is adversarially trained, this difference is much lower. We attribute this smaller decrease to the fact that adversarially trained models rely on signal features that are less susceptible to adversarial perturbations and, by extension, to small AWGN noise. Second, when comparing different testing perturbations we can see that the higher the perturbation strength the higher the fooling rate. This is intuitive since the attacker is less constrained. Third, like in the previous experiment, adversarial training always improves robustness against adversarial perturbations at the cost of accuracy on unperturbed signals. However, the stronger perturbation strength does not imply more robustness. While that is true for the VT\_CNN2\_BF model, it does not apply to the ResNet model. For high values of SPR (e.g., 15 dB), overfitting the training set can be harmful for generalization. This is specially prevalent when using more complex models like the ResNet. Thus, the perturbation strength should be treated as an hyperparameter and tuned accordingly.

\begin{table*}[htbp]
	\centering
	\small
	\parbox{.48\linewidth}{
	\vspace{.04in}
	\begin{tabular}{c|c|cccc}
	    \multicolumn{2}{c|}{} & \multicolumn{4}{c}{Test attack SPR} \\
	    \cline{3-6}
		\multicolumn{2}{c|}{} & Natural & 25dB & 20dB & 15dB \\
		\hline
		\multirow{3}{2.4em}{Train attack SPR} & Natural & $\textbf{0.452}$	& $0.099$ & $0.051$	& $0.020$ \\
		& 20dB & $0.372$	& $0.292$ & $0.251$	& $0.137$ \\
		& 15dB & $0.324$	& $\textbf{0.304}$ & $\textbf{0.278}$	& $\textbf{0.214}$ \\
    \end{tabular}
    }
    \parbox{.48\linewidth}{
    \vspace{.04in}
	\begin{tabular}{c|c|cccc}
	    \multicolumn{2}{c|}{} & \multicolumn{4}{c}{Test attack SPR} \\
	    \cline{3-6}
		\multicolumn{2}{c|}{} & Natural & 25dB & 20dB & 15dB \\
		\hline
		\multirow{3}{2.4em}{Train attack SPR} & Natural & $\textbf{0.452}$	& $0.131$ & $0.069$	& $0.033$ \\
		& 20dB & $0.372$	& $0.303$ & $0.268$	& $0.159$ \\
		& 15dB & $0.324$	& $\textbf{0.308}$ & $\textbf{0.288}$	& $\textbf{0.227}$
    \end{tabular}
    }
    \caption{Robustness (left) and security (right) results of the VT\_CNN2\_BF model for the RML2018 dataset when trained and tested with PGA-7 $l_{\infty}$ and PGA-20 $l_{\infty}$ attacks of differing SPR, respectively. ``Natural" denotes that no attack was used.}
    \label{tab:vt_cnn}
\end{table*}

\begin{table*}[htbp]
	\centering
	\small
	\parbox{.48\linewidth}{
	\begin{tabular}{c|c|cccc}
	    \multicolumn{2}{c|}{} & \multicolumn{4}{c}{Test attack SPR} \\
	    \cline{3-6}
		\multicolumn{2}{c|}{} & Natural & 25dB & 20dB & 15dB \\
		\hline
		\multirow{3}{2.6em}{Train attack SPR} & Natural & $\textbf{0.608}$	& $0.246$ & $0.177$	& $0.106$ \\
		& 20dB & $0.361$	& $\textbf{0.342}$ & $\textbf{0.325}$	& $\textbf{0.286}$ \\
		& 15dB & $0.312$	& $0.305$ & $0.300$	& $0.281$ \\
    \end{tabular}
    }
    \parbox{.48\linewidth}{
	\begin{tabular}{c|c|cccc}
	    \multicolumn{2}{c|}{} & \multicolumn{4}{c}{Test attack SPR} \\
	    \cline{3-6}
		\multicolumn{2}{c|}{} & Natural & 25dB & 20dB & 15dB \\
		\hline
		\multirow{3}{2.6em}{Train attack SPR} & Natural & $\textbf{0.608}$	& $\textbf{0.341}$ & $0.251$	& $0.154$ \\
		& 20dB & $0.361$	& $0.340$ & $\textbf{0.328}$	& $\textbf{0.293}$ \\
		& 15dB & $0.312$	& $0.301$ & $0.296$	& $0.281$ \\
    \end{tabular}
    }
    \caption{Robustness (left) and security (right) results of the ResNet model for the RML2018 dataset when trained and tested with PGA-7 $l_{\infty}$ and PGA-20 $l_{\infty}$ attacks of differing SPR, respectively. ``Natural" denotes that no attack was used.}
    \label{tab:resnet}
\end{table*}

\section{Feature analysis}
\label{sec:features}

We now compare the features learned by standardly and adversarially trained models and show that the later models have features that are more aligned with the signal statistics used by maximum likelihood methods. This insight motivates to use robust models since maximum likelihood methods are actually Bayes-optimal models for modulation recognition and could serve as reference. To understand what are the features that mainly drive the prediction of the model, we observe how the signal changes when applying adversarial perturbations. This approach is supported by several works in other application domains like computer vision \cite{Ilyas_Santurkar_Tsipras_Engstrom_Tran_Madry_2019,Engstrom_Ilyas_Santurkar_Tsipras_Tran_Madry_2019,Ortiz-Jimenez_Modas_Moosavi-Dezfooli_Frossard_2020}. Ideally, we want the model to only learn ``robust features", which are aligned with the true predictors of the task and cannot be exploited by adversarial perturbations. While ``non-robust features" generalize well and are easier to learn, since they are small-norm, adversarial perturbations can exploit them to fool the network, because their correlation with the label can be easily flipped \cite{Ilyas_Santurkar_Tsipras_Engstrom_Tran_Madry_2019}.

For our specific task, feature analysis is specially insightful. Unlike image classification, modulation recognition can be solved mathematically using maximum likelihood. This leads to the optimal model when we have perfect knowledge of the signal corruption statistics. This method gives us the exact probability for a signal to be one particular modulation and the shortest path to changing the true class of an image. Thus, if we want our models to behave like maximum likelihood methods, their adversarial perturbations should perturb the signals towards the closest theoretical boundary between classes. To simplify the analysis, we used targeted adversarial perturbations, which instead of shifting the signal towards the closest class, perturbs it towards a chosen target modulation. If we consider only AWGN noise, the optimal targeted perturbation given by maximum likelihood shifts each symbol to the closest state in the target modulation. This symbol shift can be best visualized using constellation diagrams \cite{andrew1996tanenbaum}, which we will use to verify if the model is learning meaningful features. If that is the case, the crafted perturbation should be quite aligned with the optimal one.

\begin{figure}[htbp]
    \captionsetup[subfigure]{aboveskip=-1pt}
    \centering
    \begin{subfigure}[b]{0.48\linewidth}
        \centering
        \includegraphics[width=\textwidth]{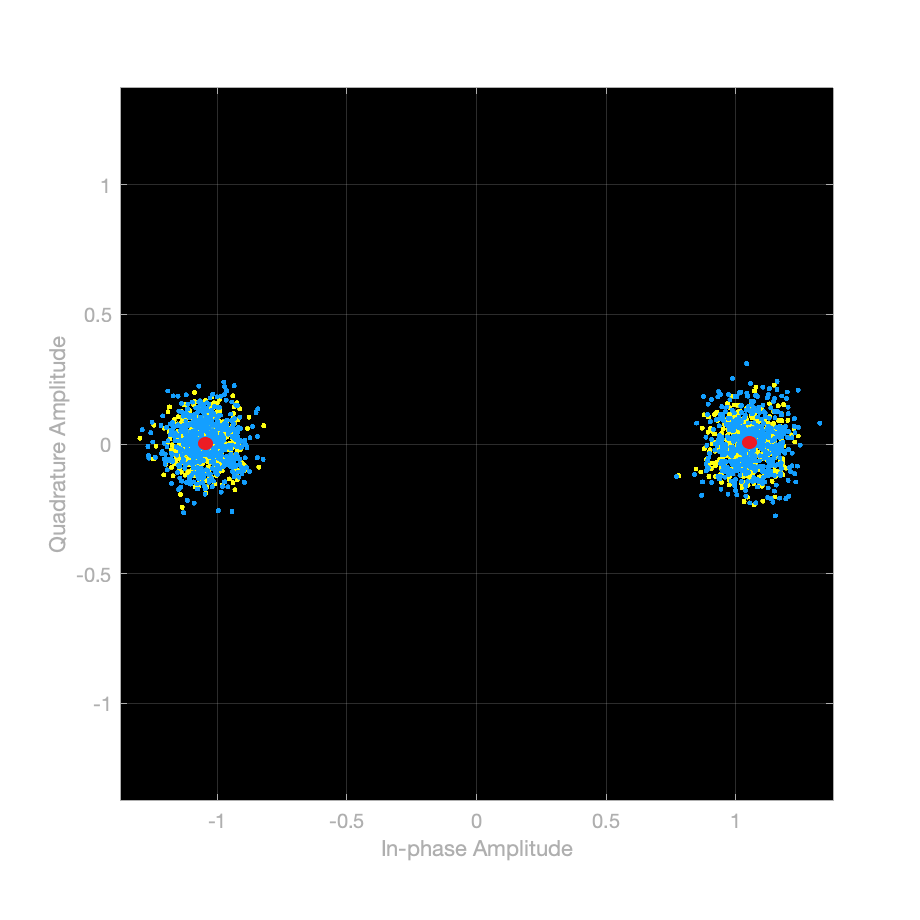}
        \subcaption{BPSK. Std. trained.}
        \label{fig:cd_2s}
    \end{subfigure}
    \hfill
    \begin{subfigure}[b]{0.48\linewidth}
        \centering
        \includegraphics[width=\textwidth]{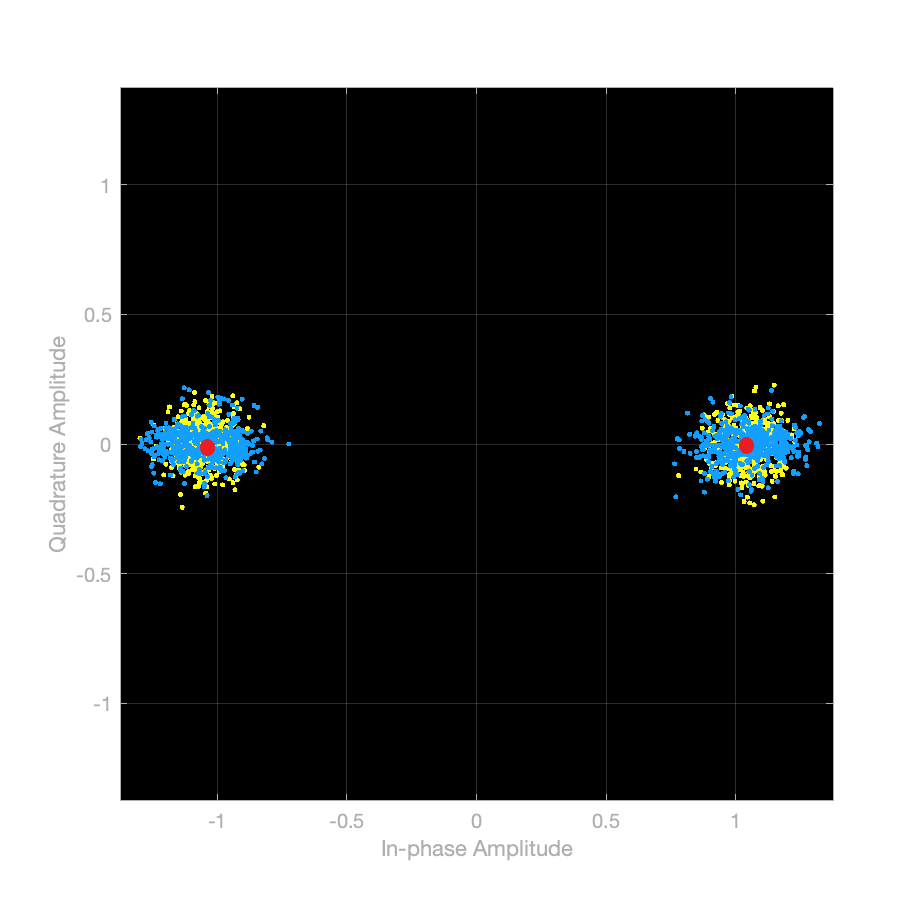}
        \subcaption{BPSK. Adv. trained.}
        \label{fig:cd_2a}
    \end{subfigure}
    \\
    \begin{subfigure}[b]{0.48\linewidth}
        \centering
        \includegraphics[width=\textwidth]{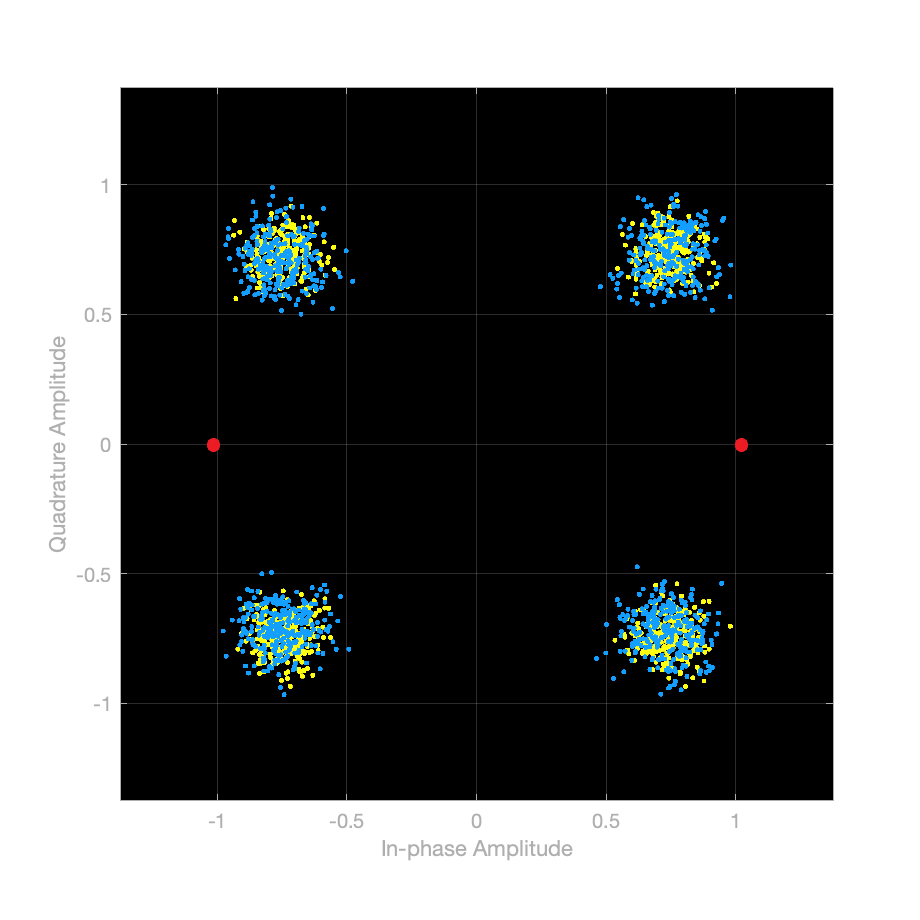}
        \subcaption{QPSK. Std. trained.}
        \label{fig:cd_4s}
    \end{subfigure}
    \hfill
    \begin{subfigure}[b]{0.48\linewidth}
        \centering
        \includegraphics[width=\textwidth]{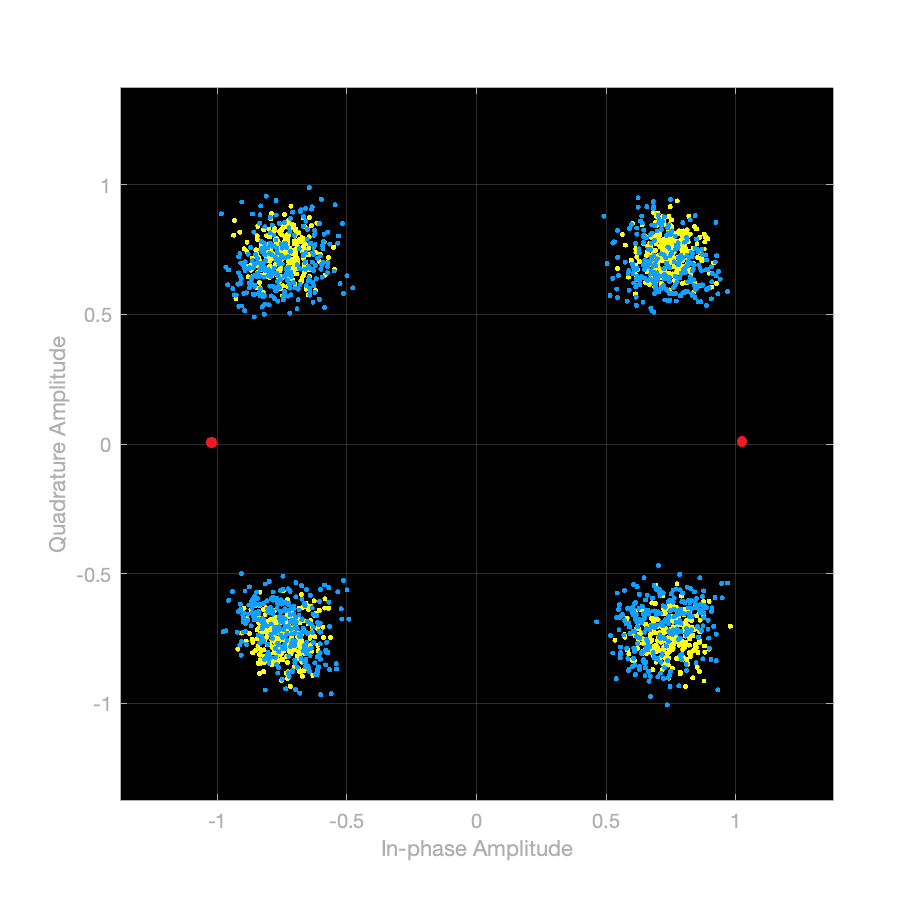}
        \subcaption{QPSK. Adv. trained.}
        \label{fig:cd_4a}
    \end{subfigure}
    \caption{Constellation diagrams of both original (yellow) and perturbed (blue) BPSK and QPSK signals. The perturbation maximizes the probability of the signal being BPSK for both a standardly and adversarially trained VT\_CNN2\_BF model.}
    \label{fig:cd}
\end{figure}

To test our hypothesis that, for robust models, the crafted adversarial perturbations should be aligned with the Bayes-optimal perturbation, we trained a VT\_CNN2\_BF model on our CRML2018 dataset and generated targeted adversarial examples towards BPSK using FGSM. By using BPSK as a target, we can see how the model thinks we should perturb the signal such that the probability of the signal being BPSK is maximized. Moreover, because in our custom dataset we only consider Gaussian noise, we know that, if the features learned are robust, like the ones used by maximum likelihood, the perturbation should shift the symbols towards the BPSK valid states. We show in Figure \ref{fig:cd} constellation diagrams where both the original and the perturbed symbols are superposed in different colors. For the plots, we show two signals from the test set: a BPSK and a QPSK signal. In Subfigures \ref{fig:cd_2s} and \ref{fig:cd_4s}, the perturbations were generated by a standardly trained model, while in Subfigures \ref{fig:cd_2a} and \ref{fig:cd_4a} we use an adversarially trained model instead. In all the plots, the perturbation has the same strength: 20 dB SPR. The red points mark the valid states (without noise) of the target BPSK modulation.

When comparing standard and adversarial training for both signals, it is clear that adversarial training makes the model learn features more aligned with the Bayes-optimal model. Although the perturbation strength is the same in all plots, the symbol shift towards the valid states of the BPSK modulation is bigger for the adversarially trained model. 
First, let us look at the generated BPSK signal. The perturbation of the standardly trained model shifts the symbols seemingly in random directions, not reducing the energy of the AWGN corruption but increasing the model confidence on the BPSK prediction. In contrast, the perturbation of the adversarially trained model is more confident after reducing the variance of the quadrature component. However, one would expect that in-phase component variance should be reduced too. We believe the reason this is not the case is the dataset class unbalance. In our dataset only the BPSK and PAM4 signals have almost no quadrature component, making it much more important than the in-phase component for prediction. 
Finally, let us look at the generated QPSK signal. When comparing the two perturbations, the shift towards BPSK is bigger in the adversarially trained model. However, the adversarial model is not fooled into predicting it as BPSK, while the standard model is fooled by the perturbation with smaller shift. 

To sum up the results of the analysis, the adversarially trained model is not only less susceptible to adversarial attacks, but it also learns better features that are clearly more aligned with the task. This can be especially useful when we want to deploy our model on unknown communication channel conditions, where there is a distribution shift between the training and test data. We expect robust features to transfer better on varying channel conditions, reducing the decrease in performance.

\section{Conclusion and Future work}
\label{sec:conclusion}

In this work, we propose adversarial training to make AMC models more robust against adversarial perturbations. We show that adversarial training greatly increases the robustness of the models and makes them less susceptible to adversarial attacks in real scenarios. Furthermore, we look at the adversarial perturbations in the constellation space to discern the features learned by the model. We show that robust models learn better features for modulation classification. That is, they correlate more with the features used by the Bayes-optimal Maximum Likelihood model. This can be measured by comparing the model adversarial perturbations with respect to their theoretical counterparts. Thus, robustness is the key to ensuring that DNNs learn features that are well-aligned with the task, and possibly transfer better between different channel conditions.

In the future, on the one hand, our analysis could be expanded beyond small $l_{\infty}$ norm adversarial perturbations to other types of corruptions that are common in AMC. For example, it would be useful to evaluate the model robustness and security to adversarial perturbations based on Rayleigh or Rician channel corruptions, which would have non-uniform constraints. At the same time, it would test if robust features transfer between channel conditions. On the other hand, from our first experiment it seems the adversarial perturbations can completely change the modulation of high-order modulations, hurting the model performance. We believe that a fairer robustness framework should have the SPR of the perturbation be dependent on the modulation of the signal it is applied to and the set of modulations the model is able to classify.

\bibliographystyle{IEEEbib}
\bibliography{references}

\begin{thebibliography}{10}

\bibitem{goodfellow2016deep}
Ian Goodfellow, Yoshua Bengio, Aaron Courville, and Yoshua Bengio,
\newblock {\em Deep learning}, vol.~1,
\newblock MIT press Cambridge, 2016.

\bibitem{Dobre_Abdi_Bar-Ness_Su_2007}
O.~A. Dobre, A.~Abdi, Y.~Bar-Ness, and W.~Su,
\newblock ``Survey of automatic modulation classification techniques: classical
  approaches and new trends,''
\newblock {\em IET Communications}, vol. 1, no. 2, pp. 137–156, Apr 2007.

\bibitem{Sun_Chen_Shi_Hong_Fu_Sidiropoulos_2017}
H.~Sun, X.~Chen, Q.~Shi, M.~Hong, X.~Fu, and N.~D. Sidiropoulos,
\newblock ``Learning to optimize: Training deep neural networks for wireless
  resource management,''
\newblock in {\em 2017 IEEE 18th International Workshop on Signal Processing
  Advances in Wireless Communications (SPAWC)}, Jul 2017, p. 1–6.

\bibitem{Chalapathy_Chawla_2019}
Raghavendra Chalapathy and Sanjay Chawla,
\newblock ``Deep learning for anomaly detection: A survey,''
\newblock {\em arXiv:1901.03407 [cs, stat]}, Jan 2019,
\newblock arXiv: 1901.03407.

\bibitem{OShea_Roy_Clancy_2018}
Timothy~J. O’Shea, Tamoghna Roy, and T.~Charles Clancy,
\newblock ``Over the air deep learning based radio signal classification,''
\newblock {\em IEEE Journal of Selected Topics in Signal Processing}, vol. 12,
  no. 1, pp. 168–179, Feb 2018,
\newblock arXiv: 1712.04578.

\bibitem{Szegedy_Zaremba_Sutskever_Bruna_Erhan_Goodfellow_Fergus_2014}
Christian Szegedy, Wojciech Zaremba, Ilya Sutskever, Joan Bruna, Dumitru Erhan,
  Ian Goodfellow, and Rob Fergus,
\newblock ``Intriguing properties of neural networks,''
\newblock {\em arXiv:1312.6199 [cs]}, Feb 2014,
\newblock arXiv: 1312.6199.

\bibitem{Moosavi-Dezfooli_Fawzi_Fawzi_Frossard_2017}
Seyed-Mohsen Moosavi-Dezfooli, Alhussein Fawzi, Omar Fawzi, and Pascal
  Frossard,
\newblock ``Universal adversarial perturbations,''
\newblock {\em arXiv:1610.08401 [cs, stat]}, Mar 2017,
\newblock arXiv: 1610.08401.

\bibitem{Sadeghi_Larsson_2019}
M.~Sadeghi and E.~G. Larsson,
\newblock ``Adversarial attacks on deep-learning based radio signal
  classification,''
\newblock {\em IEEE Wireless Communications Letters}, vol. 8, no. 1, pp.
  213–216, Feb 2019.

\bibitem{Lin_Zhao_2020}
Y.~{Lin}, H.~{Zhao}, Y.~{Tu}, S.~{Mao}, and Z.~{Dou},
\newblock ``Threats of adversarial attacks in dnn-based modulation
  recognition,''
\newblock in {\em IEEE INFOCOM 2020 - IEEE Conference on Computer
  Communications}, 2020, pp. 2469--2478.

\bibitem{Flowers_Buehrer_Headley_2019}
Bryse Flowers, R.~Michael Buehrer, and William~C. Headley,
\newblock ``Evaluating adversarial evasion attacks in the context of wireless
  communications,''
\newblock {\em arXiv:1903.01563 [cs, eess, stat]}, Mar 2019,
\newblock arXiv: 1903.01563.

\bibitem{maroto2021benefits}
Javier Maroto, Gérôme Bovet, and Pascal Frossard,
\newblock ``On the benefits of robust models in modulation recognition,''
\newblock .

\bibitem{Chung-Yu_Huan_Polydoros_1995}
Chung-Yu Huan and A.~Polydoros,
\newblock ``Likelihood methods for mpsk modulation classification,''
\newblock {\em IEEE Transactions on Communications}, vol. 43, no. 2/3/4, pp.
  1493–1504, Feb 1995.

\bibitem{Hameed_Dobre_Popescu_2009}
F.~Hameed, O.~A. Dobre, and D.~C. Popescu,
\newblock ``On the likelihood-based approach to modulation classification,''
\newblock {\em IEEE Transactions on Wireless Communications}, vol. 8, no. 12,
  pp. 5884–5892, Dec 2009.

\bibitem{Rajendran_Meert_Giustiniano_Lenders_Pollin_2018}
Sreeraj Rajendran, Wannes Meert, Domenico Giustiniano, Vincent Lenders, and
  Sofie Pollin,
\newblock ``Deep learning models for wireless signal classification with
  distributed low-cost spectrum sensors,''
\newblock {\em IEEE Transactions on Cognitive Communications and Networking},
  vol. 4, no. 3, pp. 433–445, Sep 2018.

\bibitem{Guo_Jiang_Wu_Zhou_2020}
Youwei Guo, Hongyu Jiang, Jing Wu, and Jie Zhou,
\newblock ``Open set modulation recognition based on dual-channel lstm model,''
\newblock {\em arXiv:2002.12037 [eess]}, Feb 2020,
\newblock arXiv: 2002.12037.

\bibitem{Hochreiter_Schmidhuber_1997}
Sepp Hochreiter and Jürgen Schmidhuber,
\newblock ``Long short-term memory,''
\newblock {\em Neural Computation}, vol. 9, no. 8, pp. 1735–1780, Nov 1997.

\bibitem{OShea_Corgan_Clancy_2016}
Timothy~J. O’Shea, Johnathan Corgan, and T.~Charles Clancy,
\newblock ``Convolutional radio modulation recognition networks,''
\newblock in {\em Engineering Applications of Neural Networks}, Chrisina Jayne
  and Lazaros Iliadis, Eds. 2016, Communications in Computer and Information
  Science, p. 213–226, Springer International Publishing.

\bibitem{West_OShea_2017}
N.~E. West and T.~O’Shea,
\newblock ``Deep architectures for modulation recognition,''
\newblock in {\em 2017 IEEE International Symposium on Dynamic Spectrum Access
  Networks (DySPAN)}, Mar 2017, p. 1–6.

\bibitem{Krizhevsky_Sutskever_Hinton_2017}
Alex Krizhevsky, Ilya Sutskever, and Geoffrey~E. Hinton,
\newblock ``Imagenet classification with deep convolutional neural networks,''
\newblock {\em Communications of the ACM}, vol. 60, no. 6, pp. 84–90, May
  2017.

\bibitem{Szegedy_Ioffe_Vanhoucke_Alemi_2016}
Christian Szegedy, Sergey Ioffe, Vincent Vanhoucke, and Alex Alemi,
\newblock ``Inception-v4, inception-resnet and the impact of residual
  connections on learning,''
\newblock {\em arXiv:1602.07261 [cs]}, Aug 2016,
\newblock arXiv: 1602.07261.

\bibitem{Engstrom_Ilyas_Santurkar_Tsipras_Tran_Madry_2019}
Logan Engstrom, Andrew Ilyas, Shibani Santurkar, Dimitris Tsipras, Brandon
  Tran, and Aleksander Madry,
\newblock ``Adversarial robustness as a prior for learned representations,''
\newblock {\em arXiv:1906.00945 [cs, stat]}, Sep 2019,
\newblock arXiv: 1906.00945.

\bibitem{ilyas2019adversarial}
Andrew Ilyas, Shibani Santurkar, Dimitris Tsipras, Logan Engstrom, Brandon
  Tran, and Aleksander Madry,
\newblock ``Adversarial examples are not bugs, they are features,''
\newblock in {\em NeurIPS}, 2019, pp. 125--136.

\bibitem{Cohen_Rosenfeld_Kolter_2019}
Jeremy~M. Cohen, Elan Rosenfeld, and J.~Zico Kolter,
\newblock ``Certified adversarial robustness via randomized smoothing,''
\newblock {\em arXiv:1902.02918 [cs, stat]}, Jun 2019,
\newblock arXiv: 1902.02918.

\bibitem{Salman_Li_Razenshteyn_Zhang_Zhang_Bubeck_Yang_2019}
Hadi Salman, Jerry Li, Ilya Razenshteyn, Pengchuan Zhang, Huan Zhang, Sebastien
  Bubeck, and Greg Yang,
\newblock ``Provably robust deep learning via adversarially trained smoothed
  classifiers,''
\newblock {\em Advances in Neural Information Processing Systems}, vol. 32, pp.
  11292–11303, 2019.

\bibitem{Moosavi-Dezfooli_Fawzi_Uesato_Frossard_2019}
Seyed-Mohsen Moosavi-Dezfooli, Alhussein Fawzi, Jonathan Uesato, and Pascal
  Frossard,
\newblock ``Robustness via curvature regularization, and vice versa,''
\newblock in {\em 2019 IEEE/CVF Conference on Computer Vision and Pattern
  Recognition (CVPR)}. Jun 2019, p. 9070–9078, IEEE.

\bibitem{Jagatap_Chowdhury_Garg_Hegde_2020}
Gauri Jagatap, Animesh~Basak Chowdhury, Siddharth Garg, and Chinmay Hegde,
\newblock ``Adversarially robust learning via entropic regularization,''
\newblock {\em arXiv:2008.12338 [cs, stat]}, Aug 2020,
\newblock arXiv: 2008.12338.

\bibitem{Madry_Makelov_Schmidt_Tsipras_Vladu_2019}
Aleksander Madry, Aleksandar Makelov, Ludwig Schmidt, Dimitris Tsipras, and
  Adrian Vladu,
\newblock ``Towards deep learning models resistant to adversarial attacks,''
\newblock {\em arXiv:1706.06083 [cs, stat]}, Sep 2019,
\newblock arXiv: 1706.06083.

\bibitem{OShea_West_2016}
Timothy~J. O’Shea and Nathan West,
\newblock ``Radio machine learning dataset generation with gnu radio,''
\newblock {\em Proceedings of the GNU Radio Conference}, vol. 1, no. 11, Sep
  2016.

\bibitem{Ilyas_Santurkar_Tsipras_Engstrom_Tran_Madry_2019}
Andrew Ilyas, Shibani Santurkar, Dimitris Tsipras, Logan Engstrom, Brandon
  Tran, and Aleksander Madry,
\newblock {\em Adversarial Examples Are Not Bugs, They Are Features}, p.
  125–136,
\newblock Curran Associates, Inc., 2019.

\bibitem{Ortiz-Jimenez_Modas_Moosavi-Dezfooli_Frossard_2020}
Guillermo Ortiz-Jimenez, Apostolos Modas, Seyed-Mohsen Moosavi-Dezfooli, and
  Pascal Frossard,
\newblock ``Hold me tight! influence of discriminative features on deep network
  boundaries,''
\newblock {\em arXiv:2002.06349 [cs, stat]}, Feb 2020,
\newblock arXiv: 2002.06349.

\bibitem{andrew1996tanenbaum}
S~Andrew,
\newblock ``Tanenbaum computer networks,''
\newblock {\em Computer Networks, Englewood Cliffs}, pp. 141--148, 1996.

\end{thebibliography}

\end{document}